# "Alexa, Do You Know Anything?" The Impact of an Intelligent Assistant on Team Interactions and Creative Performance Under Time Scarcity

Sonia Jawaid Shaikh[1] & Ignacio Cruz[2]

**ABSTRACT**

Human-AI collaboration is on the rise with the deployment of AI-enabled intelligent assistants (e.g. Amazon Echo, Cortana, Siri, etc.) across organizational contexts. It is claimed that intelligent assistants can help people achieve more in less time (Personal Digital Assistant - Cortana, n.d.). However, despite the increasing presence of intelligent assistants in collaborative settings, there is a void in the literature on how the deployment of this technology intersects with time scarcity to impact team behaviors and performance. To fill this gap in the literature, we collected behavioral data from 56 teams who participated in a between-subjects 2 (Intelligent Assistant: Available vs. Not Available) x 2 (Time: Scarce vs. Not Scarce/Control) lab experiment. The results show that teams with an intelligent assistant had significantly fewer interactions between its members compared to teams without an intelligent assistant. Teams who faced time scarcity also used the intelligent assistant more often to seek its assistance during task completion compared to those in the control condition. Lastly, teams with an intelligent assistant underperformed on a creative task compared to those without the device. We discuss implications of this technology from theoretical, empirical, and practical perspectives.

*Keyword*s: human-AI collaboration, intelligent assistant, interactions, teams**,** time

## 1. INTRODUCTION

In the past few years, artificial intelligence (AI)-enabled interactive technologies such as intelligent assistants (e.g. Alexa, Cortana, Siri, etc.) have been deployed across individual and organizational contexts (e.g. space missions, companies, etc.) to assist individuals and teams with various tasks (Bogers et. al., 2019; Crook, 2017; Sentance, 2018; Wall, 2018; Wilson & Daugherty, 2018; Canbek and Mutlu 2016). These intelligent assistants run on complex algorithmic systems (e.g. Natural Language Processing), which allow them to be interactive and

[1] Annenberg School of Communication, University of Southern California
  Corresponding author: soniajas@usc.edu

[2] Annenberg School of Communication, University of Southern California

Acknowledgement: This project was funded by the Annenberg Summer Research Fellowship 2017 and partially supported by the National Science Foundation Graduate Research Fellowship No. 20162238.

assistive in nature (Hausfeld et. al., 2013; Hoy, 2018). Most commercially available intelligent assistants are conversational in nature, meaning they can understand verbal commands or questions. This allows users to delegate assignments to them without entirely disengaging from their current task (Luger & Sellen, 2016; Goksel & Mutlu, 2016; López, Quesada, & Guerrero, 2018). For instance, a team who is preparing for an important presentation may use verbal communication to ask the intelligent assistant to set project reminders or find information instead of doing it by themselves. Given the unique abilities and functions of this technology, it is understandable that intelligent assistants are touted as machines that help users do more in less time (Personal Digital Assistant - Cortana, n.d.; CIMON brings AI to international space station, n.d.). Recent estimates show that the market for intelligent assistants will reach $25.63 billion by 2025 and will continue to expand thereafter (Business Wire, 2019). Therefore, we can presume that intelligent assistants will become a part of our lives and have the enormous potential to transform the ways we collaborate, interact, and accomplish our goals (Këpuska & Bohouta, 2018; Miner et. al., 2016).

Interestingly, the rate at which we are deploying these AI-enabled intelligent assistants is much higher than the rate at which we are testing them for their effects on teams in collaborative settings, especially under time scarcity. Time is one of the most important and fundamental resources for most task-oriented teams and individuals as it impacts their productivity, and intergroup behaviors (Durham et. al., 2009). Therefore, it is crucial to understand how the AI-enabled technology such as an intelligent assistant can affect their behaviors under time scarcity. We conducted a study on 56 dyadic teams to help fill some part of the gap in the literature and to provide evidence on how the deployment of an intelligent assistant affects team members' interactions with each other, performance, and technology use. We used the meta-perspective of bounded rationality (Newell & Simon, 1972; Simon, 1955; 1972a; 1976b; 1991c)—to foreground our argument because the choice of using a technology under time scarcity fits well with this theoretical narrative that focuses on the role of environmental constraints on decision-making. Additionally, we used the recent literature from algorithm appreciation (Logg, Minson, & Moore, 2018;), and algorithm aversion (Dietvorst, Simmons, & Massey, 2015; Yeoman et. al., 2018) to develop our hypotheses.

In the following sections, we begin by describing the meta-theoretical framework of bounded rationality and how it applies to the use of intelligent assistants under time scarcity. We then focus on how an intelligent assistant can influence team members' interactions with each other. We also investigate how teams' creativity differs by time scarcity and intelligent assistants. These theoretical sections are followed by an explication of our methodology, results, and a discussion of our findings. We end with implications and directions for future research.

## 2. THEORETICAL FOUNDATION: BOUNDED RATIONALITY

The homo economicus (Latin for 'economic man') model of rationality argues that decision-makers are only motivated by their self-interests, and therefore, make decisions that can help maximize rewards or pay-offs without any consideration for others (Archer, 2013; Mill, 1863; Sen, 1977). According to this perspective, decision-makers can rank-order the alternatives (i.e. their options) in terms of utilities or usefulness, which makes it easier for them to make the 'optimal' or the best choice. Therefore, the underlying assumption of this model is that decision-makers have the cognitive capacity to process a set (or more) of alternatives to make an 'optimal' decision. Since its conceptualization in the late 19th century and further developments in the 20th century (see Oppenheimer, 2008); the homo economicus model has produced a large body of research that has tested and often falsified its basic premises and assertions across a variety of contexts (see Gintis, 2000; Henrich et. al., 2001; Kahneman & Tversky, 1979; Thaler, 2000).

One of the first comprehensive theoretical disagreements with the homo economicus model was presented by Herbert Simon in his framework called *Bounded Rationality,* which emerged from his studies in organizational behaviors (Simon 1955; March & Simon, 1958). He argued that decision-makers are often embedded in limiting environments where they may face lack of time or information overload. These constraints interact with their cognitive limitations to yield a boundedly rational form of decision-making. This suggests that a decision-maker cannot always rank order her alternatives, or compute utilities to search for the best possible alternative. Instead she tries to satisfice--i.e. engages in a process that allows her to choose what is good enough in that environment (Newell & Simon, 1972; Simon, 1991). The focus of bounded rationality perspective was not on the role of individual choices in predicting good or bad outcomes; but on the process of making choices given an environment and its parameters. As a meta-perspective of decision-making, bounded rationality explicitly factored in realistic environmental parameters and human cognitive limitations that affect a decision-maker and her choices (Simon, 1955).

In line with the bounded rationality framework, researchers have suggested that individuals and groups use heuristics (i.e. mental shortcuts) to make decisions when they find themselves under pressing conditions (Gigerenzer & Goldstein, 2011; Todd & Gigerenzer, 2000; Gigerenzer et al. 1999) such as time scarcity (Ariely & Zakay, 2001). Heuristics serve as guiding principles to make a choice which is 'good enough' rather than search for the best possible option under limiting circumstances. For instance, when people encounter or feel resource scarcity; they are more likely to feel cognitively constrained and therefore, engage and process the situation heuristically and focus on one alternative or strategy while disregarding all the available options others (Janis, 1983; Mani, Mullainathan, Shafir & Zhao, 2013; Mullainathan & Shafir, 2013; Zhu & Ratner, 2015). Consider the case of a team that is working under a looming deadline. The members of the team may decide to forgo deliberations or brainstorming amongst each other as alternatives in favor of completing the assignment as quickly as possible. This may help them complete the task on time but may result in work that could have been more creative (see Kelly & Krau; 1991).

Technologies such as intelligent assistants also function as one of the many alternatives (e.g. human labor, other types of technologies, etc.) in any collaborative setting, which means that decision-makers have the choice to use them or not. However, their deployment is often justified with the reason that they can help maximize productivity and improve outcomes (Personal Digital Assistant - Cortana, n.d., López, Quesada, & Guerrero, 2018). This implies that amongst the two alternatives i.e. using an intelligent or not, the former may be a more rational choice. We contend that it is more important to understand how contextual and environmental factors such as time scarcity *rationalize* the use of this technology rather than simply labelling this use as rational or irrational. In the following section, we delve deeper into how time scarcity as an environmental factor can increase the use of an intelligent assistant.

## 2.1. Intelligent Assistant Use and Time Scarcity

People use intelligent assistants for various reasons such as finding information (e.g. weather updates), playing music, navigation, sending messages, online shopping, etc. (Dubiel, Halvey, & Azzopardi, 2019; Sentance, 2018). However, humans' acceptance and use of AI-enabled technologies has had its ups and downs. Some research on human-AI relationship has shown that individuals often display algorithm aversion i.e. the tendency to discount advice from an algorithm. This effect often occurs when humans find an algorithmic system' performance uncanny or when it outperforms humans (Dietvorst, Simmons, & Massey, 2015; Yeoman et. al., 2018) This aversion has been found to manifest as an individual's preference for human advice over those given by AI-enabled or algorithmic technology (Prahl & Van Swol, 2017). Some reasons that researchers have given to explain algorithm aversion include humans' lower trust in automated systems, lack of anthropomorphic qualities in AI-enabled technologies, and a poor understanding of how these algorithms work (Prahl & Van Swol, 2017; Yeoman et. al., 2018). However, some research has found evidence contrary to this effect. For instance, there's evidence that individuals prefer algorithms when solving logic problems (Dijkstra, Liebrand, & Timminga, 1998; Dijkstra, 1999). Similarly, in a series of recent studies, it has been shown that people preferred advice from an algorithm to that from another human. This effect sustained even when people chose to seek advice both from an algorithm and a human as they still preferred the former's evaluations (Logg, Minson, & Moore, 2018). This effect has been termed as *algorithm appreciation* (Logg, Minson, & Moore, 2018). Some explanations for why people may prefer algorithmic systems may have to do with users' positive bias and greater trust towards trusting and relying on the technologies (Booth et. al., 2017; Clark, Robert, & Hampton, 2016). Given the state of research, Logg, Minson, & Moore (2018) correctly note that when and why humans may or may not prefer algorithmic systems is a rich and murky area that that has produced a variety of contrasting findings.

It is important here to note that most of the literature on the use of AI-enabled technology has generally focused on individual-level (e.g. advisor credibility, trust), algorithm-level (e.g. error-prone technology, lack of explainable AI), and/or task-level (e.g. logic problems) factors to explain why people may or may not use an AI-enabled technology (Dietvorst, Simmons, & Massey, 2015; Dijkstra, Liebrand, & Timminga, 1998; Prahl & Van Swol; 2017; Logg, Minson, & Moore; Yeoman et. al., 2018). However, there is a void in research on the role of environmental constraints on the preference and use of algorithmic or AI-enabled technology. Research from the domain of human-robot interaction provides some evidence that individual use and engagement with mobile intelligent assistants is amplified under environmental constraints (Robinette, Howard, & Wagner, 2017; Robinette, Li, Allen, Howard, & Wagner, 2016). For instance, lab studies show that decision-makers chose to use the advice of a robot (a mobile physical assistant) to navigate their way out of time-critical fire evacuation scenarios despite having known about a robot's poor performance via prior interactions (Robinette, Howard, & Wagner, 2017; Robinette, Li, Allen, Howard, & Wagner, 2016). These findings demonstrated algorithm appreciation, but under the condition of temporal constraint. Merging perspectives from this literature and the overarching bounded rationality framework (Simon, 1955; Todd & Gigerenzer; 2000), and scarcity (Mullainathan & Shafir, 2013), and algorithm appreciation (Logg, Minson, & Moore, 2018); similar theoretical logic can be extended to the use of intelligent assistants. Arguably, the decision to use an intelligent technology is a type of mental shortcut (see Sundar, 2008) that emerges in limiting circumstances. Therefore, we hypothesize that:

$H_1$: *Teams under time scarcity are more likely to use the intelligent assistant--i.e. they will request more assistance from an intelligent assistant--compared to the teams without time scarcity to complete the same task.*

In the above section we hypothesized that under time scarcity, teams are more likely to use the machine. However, team members need to do more than just use technology to complete a task—they also need to communicate with each other. Previously, team members' communication was studied under the umbrella of computer-mediated communication which can be defined as the communication between two or more individuals via some sort of technology (Walther, 1996). For example, emails allow teammates to communicate via technology or through a channel/medium. Yet, the inclusion of an interactive device changes how we think of group interaction. As the definition suggests, intelligent assistants are endowed with an interactive ability in addition to being task performers for humans (Hausfeld et. al., 2013). Therefore, the deployment of intelligent assistants as a technology presents a different kind of communication scenario: namely, human-machine interaction (c.f. Card, 2018). In this case, the communication between users is not being mediated by technology; but rather, their interactions are being established *with* the machine in addition to the communication between human teammates as they pursue a goal. This property not only renders intelligent assistants apart from other technologies such as emails, Skype or a social networking site, but also changes the parameters of how communication is established between users and the technology (Guzman & Lewis, 2019). Inadvertently, the deployment of an interactive technology can also impact how members communicate with each other. In the next section, we expand on how intelligent assistants can influence group interactions.

## 2.2. Team Members' Interactions and Intelligent Assistant

Team members' interactions are pivotal to their group's development, survival, performance and goal accomplishment (Mesmer-Magnus & Dechurch; 2009; Johnson et. al. 2002; Potter et al. 2000). Team members' interaction with each other allow them to exchange ideas and information, create and resolve conflict (Franco, Rouwette, & Korzilius, 2018), form friendships, delegate tasks, and complete projects (Merritt et. al., 2013). One factor that negatively impacts group interaction is time scarcity. For instance, groups reduce information exchange (e.g. sharing and seeking information) with others (Durham et. al., 2009) under time scarcity, making it difficult to establish consensus

The deployment of new AI-enabled technologies in teams who face time scarcity can change how members interact with each other. Although, the research is yet to catch up in this area, literature on the impact of a physical technology such as a cell phone has shown that the 'mere-presence' of a cellphone negatively influenced dyadic partners' perceptions of their conversation and relationship quality, closeness, and empathy. Further studies in this area have also found similar evidence of decreased conversational satisfaction in the presence of a cellphone (Misra, Cheng, Genevie & Yuan, 2016); Allred & Crowley, 2017; Crowley, Allred, Follon, Volkmer, 2018).

While these studies help advance our understanding concerning the effects of physical technologies on face-to-face interaction, there are some limitations. First, these studies relied exclusively on self-reported data on a participant's perception of conversation, perception of partner empathy, and relationship satisfaction. There is little or no mention of data pertaining to the actual volume of interaction exchanged between partners. Second, they also exclusively focus on the *silent* "presence" of a device. The participants were not allowed to interact with their phone, and therefore, and a lack of contextual factors that may impact quality of interaction (Crowley et. al., 2018; Miller-Ott & Kelly, 2015; Przylbyski and Weinstein, 2013).

Unlike a smartphone, an intelligent assistant cannot be described as merely present because it is essentially an interactive device which is deployed to be used by all the team members working within a collaborative setting. It can only operate if a human engages with it through verbal communication (see Cowan et. al., 2017). This means that team members must allocate a portion of their communication to an intelligent assistant and thereby decreasing interactions with each other. This effect is likely to be more pronounced under the condition of time scarcity which reduces group information exchange (Durham et. al., 2009) and possibly increase a team's use of an intelligent assistant (see hypothesis 1). Therefore, we hypothesize:

$H_2$: *Teams with an intelligent assistant available to them are more likely to have lower interpersonal verbal interactions between members compared to teams without the device. This effect will be moderated by time scarcity.*

Team members' interactions with each other directly impact their performance (Durham et. al., 2009). Hypothetically, if these machines can affect teams' interactions; then it is pertinent to explore how they influence teams' creative performance.

## 2.3. Team Creative Performance and Intelligent Assistant

Human creativity is a complex phenomenon which requires the shuffling of ideas, thinking out-of-the-box, and using various strategies to find uncommon solutions. Creative performance can be affected by many factors that are personal, social, motivational and organizational in nature (Baer, Oldham, Jacobsohn & Hollingshead, 2008; Simonton, 1999; Zhang, Tsui, & Wang, 2011). One factor that negatively impacts team creativity is time scarcity--a condition which produces cognitive constraint. Teams working under less time produce solutions lower in creativity compared to teams who have more time (Amabile, Hadley & Kramer, 2002; Baer & Oldham, 2006; Kelly & Karau, 1991). This is in line with much research that has found that time pressures negatively affect performance and strain individual and team outcomes across many contexts (Caballer, Gracia, & Peiró, 2005; Cannon-Bowers & Salas, 1998; De Dreu, 2003; Driskell, Salas, & Johnston, 1999; Ellis, 2006; Hsu & Fan, 2010). The pressure to accomplish much with limited resources makes it more difficult for team members to come to a consensus about the task at hand, which hinders the amount of communication and information exchange within the team (Cannon-Bowers & Salas, 1998). This causes the team members to reduce their attention spans and begin to be self-focused instead of actively contributing to the team (Driskell et al., 1999), and thus impacting various collaborative outcomes including creativity (Kelly & Karau, 1991; Shremata, 2000).

There is some research on how intelligent technologies impact performance for certain types of tasks in individual contexts. For instance, Edwards et. al. showed that interaction with an intelligent assistant decreases users' performance on a cognitively demanding language-based task which require content generation compared to a less stressful copying task (2019). Similarly, research on other types of AI-enabled technologies such as smartphones has shown that its presence reduces individuals' cognitive capacity and thereby negatively affecting logical and creative problem solving (Ward, Duke, Gneezy & Bos, 2017). Furthermore, there is evidence that the use of AI-enabled search engine reduces the amount of information individuals hold in their short-term memory (Sparrow, Liu, & Wegner, 2011; Wegner & Ward, 2013). These studies reflect how deployment of AI-enabled technologies--albeit of different kinds--impact performance in individual contexts. Given the current state of research, it can be argued that teams with an intelligent assistant will be less creative than teams without an intelligent assistant. However, it can also be counter-argued that due to the computational capacity of an intelligent assistant; it is possible that it can search for novel information and therefore, help teams demonstrate higher creative performance under time scarcity. To investigate these competing arguments, we pose the following question:

*RQ1. How is a team's creative performance impacted by the availability of an intelligent assistant under time scarcity?*

## 3.  METHOD

### 3.1. Participants & Selection Criteria

Originally, 60 teams participated but four had to be removed due to audio recording malfunction or violation of study protocols by the participants. This study utilized a total of 112 participants, comprising 56 two-person (dyadic) teams. All participants were undergraduate students from a West Coast university. Each participant was offered a $10 Amazon gift card and course credit in exchange for their participation in the study. Participants were 88% (*N=99)* female. We believe that a higher proportion of female to male students in our degree program can explain why we had more female research participants in our study. Prior to being selected for the study, participants were screened to ensure they had no prior exposure to or experience with a smart-speaker type of an intelligent assistant (including Amazon Echo or Google Home). We used this protocol to create an equal starting point for all participants and remove any potential confounds that may be associated with prior intelligent assistant experience. Study inclusion criteria also required that the participants considered themselves to be native speakers of English language.

### 3.2. Procedure

Participants who were eligible to participate in this study worked in teams of two and were randomly assigned to a 2 (Intelligent Assistant: Available vs. Unavailable) x 2 (Time: Scarce vs. Not Scarce/Control) between-subjects factorial design. Upon arrival to the lab, participants were randomly assigned into dyads and experimental conditions, and were led to a room to be seated around a table. Participants completed a consent form and filled out a pre-study survey. The participants were then given instructions for the creativity task which required participants to imagine themselves as a part of an event planning team who were tasked with organizing a creative Sweet 16th birthday party for a valued client with no limits on the budget. We designed and implemented this task because it offered many advantages: a) participants were undergraduates in their late teens and early twenties and thus, the age difference between our participants and a hypothetical 16-year-old client was small enough to create relevance; b) the design of the task enabled participants to focus on meaningful, pertinent ideas like creative themes that majority of participants were supposed to below the drinking age, designing a party for a 16-year old would require no use of alcohol; ) birthday parties are events individuals often attend and are familiar with; and e) Party organization can involve many creative elements.

Teams were responsible for planning three main components for the party: cake, food, and entertainment. Each team was told that they would be judged by independent raters on how creative their birthday party ideas were on the three components. We also included an incentive where all the participants were told that the best rated team would win an additional Amazon gift card at the end of data collection for the study. Each team was provided one laptop where a word processing template had instructions and a space for teams to type and record their ideas. Participants were told that they were not allowed to use smartphones, or browsers on the laptop. The experimenter then proceeded to give instructions on the intelligent assistant use (see below). This was followed by instructions on time available to them (see below). Once these procedures were completed, the experimenter informed them about the audio recorder and took verbal consent (which was in addition to the written consent on the use of audio recorder). This was followed by the experimenter turning on an audio device and leaving the room. Once the allotted time was over, or teams indicated that they were done with the task; the participants took a post-study survey. This was followed by a brief interview, debriefing and giving a $10 Amazon gift card as compensation.

**3.2.1 Intelligent assistant training and instructions.** In the intelligent assistant availability conditions, participants were given the following instructions: "You are not allowed to use a laptop browser or your smartphone for this task; however, you may use Amazon Echo if you like". The intelligent assistant was already placed on the table before the participants' arrival. The instructor then demonstrated how to use the intelligent assistant which was done via the following instructions: "Here is how to use Amazon Echo. 'Alexa, what is the weather in Los Angeles?' [response from the intelligent assistant]. 'Alexa, what is the capital of China?' [wait for response from the intelligent assistant]."

**3.2.2 Time availability manipulation.** The participants were given the following instructions to manipulate time scarcity: "Normally this task takes 30 minutes, but today you have 15 minutes to complete this task." Participants in the control condition were told: "Normally, this task takes 30 minutes to complete. You also have 30 minutes to complete this task." This technique for manipulating time was adapted from Ellis (2006).

### 3.3. Measures

**3.3.1. Interactions.** Team verbal interactions were measured by totaling the amount of turns taken by each team member within a team. Two types of verbal interaction for teams with an intelligent assistant were calculated: a. all conversational turns i.e. amongst the two teammates and the intelligent assistant and b. conversational turns between the teammates only. Turn taking was categorized using a technique adapted from Sacks, Schegloff, & Jefferson (1974). The audio recorded conversations between team members were transcribed and time stamped. All speakers were assigned the names: Speaker 1, Speaker 2 or Speaker 3 (i.e. Alexa). Crosstalk or overlap were not categorized as a turn.

**3.3.2. Use of the intelligent assistant.** The intelligent assistant used in this study, Amazon Echo, can be used via a wake word, 'Alexa'. We counted the number of times the word 'Alexa' was spoken. 'Alexa' was only included in the data analysis when it preceded a question, command, or task delegated by the team. Therefore, we did not count the word 'Alexa' in instances where team members may just be talking *about* the machine rather than to it.

**3.3.3. Team creativity.** The creative task required participants to design a Sweet 16 birthday party which included three main components: cake, food and entertainment. To measure teams' creative performance, we developed a scoring rubric with the assistance of several raters who were blind to the study hypotheses and methods. In the first step of creating this rubric, we recruited two raters who were given the following instructions: "We have documents that have creative ideas on planning a Sweet 16[th] birthday party. We need help with developing a rubric than can help people evaluate these ideas. An ideal rubric will help with scoring teams on the quality of creative ideas." To create this rubric, the individuals were provided with a randomly sampled selection of teams' birthday party ideas. Once a rubric was developed, this rubric was then passed to another blind rater who used it on a different sample of documents and helped identify any errors and make suggestions. The final rubric was then used by two independent raters who applied it to cake, food, and entertainment ideas. These raters assigned grades (A, B, C, and D which were based on the rubric) to these categories. We added the ratings across the categories to create a total creativity score for each team. The grades were converted to A = 4, B = 3, C = 2, and D = 1. Each team could attain maximum points of 12 and a minimum point of 0. Since all teams contributed to at least one category of party planning, therefore, there were no two zeros. Intercoder reliability between raters was measured via Krippendorf's alpha which was found to be 0.93.

## 4. RESULTS

We first begin by conducting a manipulation check for time availability. All individuals responded to: *Did you think you needed more time completing the task?* (1= *Definitely not*, 5 = *Definitely yes*) in a post-study survey. We divided the groups into two: scarcity and control. A one-way ANOVA showed significant differences, omnibus $F(1, 110) = 8.971$, $p < 0.05$. Groups under time scarcity indicated they needed more time to complete the task ($M = 3.09$, $SD = 1.37$) compared to groups who were not under time scarcity ($M = 2.29$, $SD = 1.43$).

Hypothesis 1 predicted that under time scarcity, teams will use the intelligent assistant more often than under the condition than those without time scarcity. To test the hypothesis, we analyzed the data in multiple ways because the two groups (scarcity and control) had unequal amount of time (15 minutes vs. 30 minutes) available to them to complete the task. We began by comparing only the first fifteen minutes of conversation within teams for both the groups and the total number of times they used or asked for help from the assistant. We counted the number of times the word "Alexa" was spoken. We performed a t-test to see how the two independent groups varying on time (scarcity vs. control) used the intelligent assistant for the first fifteen minutes. Levene's test for equality of variances was violated with $p > 0.05$, therefore Welch-Satherwaitte adjustment was applied. The resulting t-test was significant, $t = 2.67$, $p < 0.05$. In the first 15 minutes of conversation, groups with scarce time on average used the intelligent assistant more ($M = 8.35$, $SD = 5.62$) compared to groups who had more time ($M = 3.85$, $SD = 2.85$). The resulting effect size was large, $d = 1.00$ (see Cohen, 1969). These findings support our first hypothesis which predicted that teams under time scarcity are more likely to use an intelligent assistant compared to teams who do not face temporal constraints.

Our second set of analyses accounted for the entire time that was available to the teams. We compared the total number of times the intelligent assistant was used across the entire allotted time given to the two groups i.e. 15 minutes and 30 minutes. The resulting t-test was significant ($t = 2.17$, $p < 0.05$, df = 54). The mean number of times the intelligent assistant was utilized was higher in 15 minutes of conversation ($M = 8.35$, $SD = 5.62$) than 30 minutes of conversation ($M = 5.43$, $SD = 4.34$), yielding a medium effect size of $d = 0.58$.

A post-hoc analysis was performed to show how the team's use of the intelligent assistant was distributed by the time interval of five minutes. Results showed that teams under time scarcity conditions on average surged in their use of the intelligent assistant after the five-minute mark (see Figure 1). However, teams in the control condition i.e. teams without any temporal constraint, showed a gradual decline over five--minute intervals in their use of the intelligent assistant.

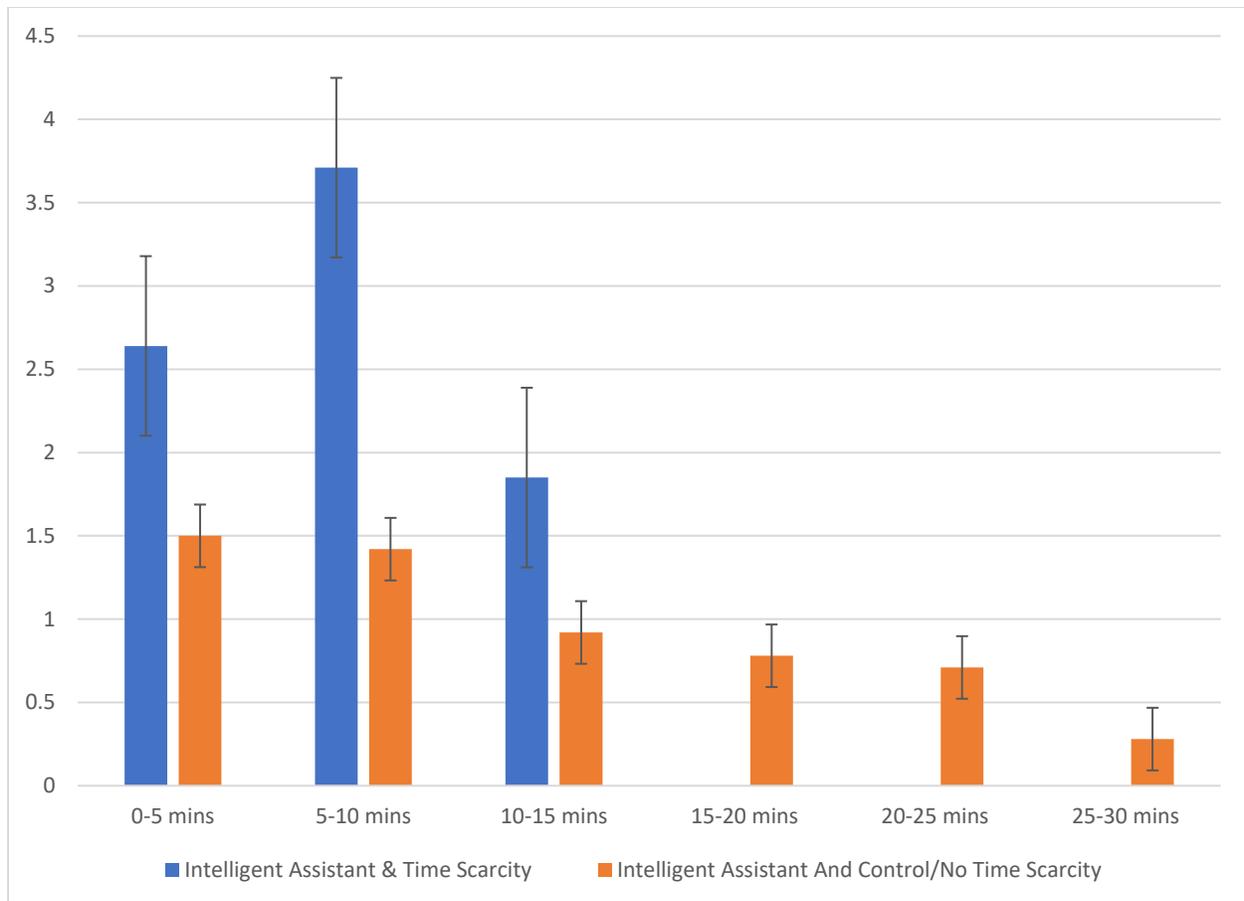

*Figure 1. Average number of times teams called "Alexa" during allotted tasked time distributed by five minutes intervals. N = 28 or n = 14 per condition.*

*Table 1.* Average number of verbal turns in each condition of the experiment across the entire time span of the task where time scarcity =15 mins, no time scarcity/control = 30 minutes).

| Conditions | Intelligent Assistant Available *M* (*SD*) | Intelligent Assistant Unavailable *M* (*SD*) | Total *M* (*SD*) |
|---|---|---|---|
| Time Scarcity | 196. 21 (86.60) | 241.35 (68.01) | 218.78 (79.79) |
| Control | 323.07 (140.59) | 418.64 (154.97) | 370.85 (153.13) |
| Total *M* (*SD*) | 259.64 (131.53) | 330.00 (148.12) | 294.82 (143.26) |

*Table 2*. Creativity scores in each condition of the experiment.

| Conditions | Intelligent Assistant Available *M* (*SD*) | Intelligent Assistant Unavailable *M* (*SD*) | Total *M* (*SD*) |
|---|---|---|---|
| Time Scarcity | 6.14 (2.98) | 7.28 (2.84) | 6.71 (2.91) |
| Control | 6.14 (2.47) | 8.85 (2.28) | 7.50 (2.71) |
| Total *M* (*SD*) | 6.14 (2.69) | 8.07 (2.65) | 7.10 (2.81) |

Hypothesis 2 predicted that with an intelligent assistant available to them, teams are more likely to have lower verbal interaction compared to teams without the device. This hypothesis was tested also tested in two ways: including and excluding team interactions with the assistant. First, we used the total number of turns taken by team members when communicating with only each other as a measure of interpersonal interaction for each team. A two-way ANOVA yielded a significant main effect of intelligent assistant $F_{(1, 52)} = 4.95$, $p < .05$, $\eta_{p^2} = 0.08$. The mean number of interactions for teams without an intelligent assistant was greater ($M = 330.00$, $SD = 148.12$) than for teams with an intelligent assistant ($M = 259.64$, $SD = 131.53$). The main effect of time availability yielded an F ratio of $F_{(1, 52)} = 23.16$, $p < .001$, $\eta_{p^2} = 0.30$ indicating that the mean number of interactions in teams under time scarcity condition was lower ($M = 218.78$, $SD = 79.79$) compared to teams who were in the control condition i.e. without time scarcity ($M = 370.85.53$, $SD = 153.13$), $d = 1.24$—also shown Table 1. The interaction effect was non-significant, $F_{(1, 52)} = 0.63$, $p > .05$.

Another set of analyses included using the total number of interactions between the speakers. This analysis held up our previous findings and found that the mean number of interactions for teams without an intelligent assistant available to them was still greater ($M = 330.00$, $SD = 148.12$) than for teams with an intelligent assistant ($M = 268.00$, $SD = 132.30$), with an effect size of $d = 0.44$. A two-way ANOVA a yielded a main effect of time availability on interactions with an F ratio of $F_{(1, 52)} = 23.64$, $p < .001$, $\eta_{p^2} = 0.31$ and a marginally significant effect of intelligent assistant availability on interactions, $F_{(1, 52)} = 3.85$, $p = 0.05$, $\eta_{p^2} = 0.06$.

To investigate how a team's creative performance is impacted by the availability of an intelligent assistant, we ran a two-way ANOVA using the composite creativity score as a dependent variable for each team. Interestingly, there was a significant main effect of intelligent assistant availability $F(1, 52) = 7.35 \ p < .05$, $\eta_{p^2} = 0.124$ such that that the mean creativity score for teams without an intelligent assistant was greater ($M = 8.07$, $SD = 2.65$) than for teams with an intelligent assistant ($M = 6.14$, $SD = 2.69$). There was no significant main effect of time availability $F(1, 52) = 1.22$, $p > .05$ or an interaction effect $F(1, 52) = 1.22$ , $p > .05$. However, although not significant, creativity score was the lowest under time scarcity ($M = 6.71$, $SD = 2.91$) compared to the control condition ($M = 7.50$, $SD = 2.71$, $d = 0.28$; see Ta

## 5. DISCUSSION

The rise of artificial intelligence (AI)-enabled technologies across homes and organizations has ushered into the fourth industrial revolution where humans and machines are collaborating from simple to complex tasks (Wall 2018, Dauherty & Wilson,). AI-enabled intelligent assistants become more relevant and casual fixtures in our personal, work and virtual spaces (see Hill, Ford, & Farreras, 2015). More often than not, these AI-enabled technologies are deployed under environments where decision-makers face time scarcity (Personal Digital Assistant - Cortana, n.d..; Canbek and Mutlu 2016). It is argued that these technologies facilitate our lives and help users who face environmental constraints (Clark et al., 2016). Some examples of such interactions and decisions include using machine-generated recommendations to choose one restaurant over another, using app-based maps to navigate and reduce travel time, and relying on algorithms to order online versus in store (e.g. Kleinberg, 2018). Therefore, the interplay of time and technology permeates the fabric of our lives it is only pertinent to understand how they impact our behaviors and performance. In this study, we have provided empirical evidence of how the availability of an intelligent assistant impacts verbal interaction, decision-making, and creativity within teams who faced temporal constraints. However, we contextualized our study for a specific outcome which is a creative performance.

We had hypothesized that teams working under time scarcity would be more likely to use or rely on the intelligent assistant. In line with this hypothesis, we found that teams under time scarcity utilized the intelligent assistant much more than teams without time scarcity (see Figure 1). We equalized the teams for a fair comparison by dividing conversation patterns by time periods. Under time scarcity, teams appear used the intelligent assistant more often to complete their tasks. Interestingly, when we included intelligent assistant use across entire time spans allotted to the teams, those under a condition of time scarcity (i.e. 15 minutes) used the device more on average compared to teams in the control condition (i.e. 30 minutes). Additionally, teams under time scarcity increased their use of the intelligent after the five-minute mark. This implies that as the time became scarcer, teams heuristically increase their use of the device. Furthermore, we found that while teams under time scarcity kept resorting to seeking assistance from an intelligent assistant; teams in the control condition showed a gradual decrease in their dependence on the intelligent assistant (see Figure 1). These changes and differences in groups' technological behaviors with respect to decreasing time are also indicative of how groups morph and develop in 'punctuated equilibriums' (Gersick, 1988, pp. 16). Groups have been found to approach and complete tasks in nonlinear development stages where some a phase of inertia is followed by vigorous activity, which often occurs as members realize decreasing time (Gersick, 1988; 1991a). Overall, these findings are consistent with the larger bounded rationality framework and the research findings from human-robot interaction where individuals tend to increase the use the technology as environmental constraints (Robinette et. al., 2017; Robinette et. al., 2016). This is possibly adding to the evidence that the use of AI-enabled technology emerges as a heuristic under environmental constraints.

The natural question which follows the above observation is: what it about AI-enabled technology that attracts users to seek help from it under environmental constraints? A possible explanation is that AI-enabled technologies are often perceived and marketed as intelligent, clever, and computationally sophisticated (Vincent, 2019). These qualities act as a signal that AI is dependable and can be trusted. Therefore, it is understandable that when faced with time constraints, decision-makers look up to this technology for help and use it more often. This is akin to human behaviors and preferences for individuals higher in authority, intelligence, and strength when resources in the environment are limited or unequally distributed (Sprong et. al., 2019)

In case of verbal interactions, we found teams who had an intelligent assistant available to them interact less with each other even when they had abundant time to complete a task compared to groups without the device. This effect is supported even when interactions with the device are included in the data analysis. Based on our available data, we suggest that the reduction of this interaction in the presence of the device may have led the participants to rely less on one another and more on the intelligent assistant —particularly under time scarcity. This is potentially a serious consequence for knowledge, information exchange and brainstorming in working groups.

At this point we must ask how the deployment of an intelligent assistant under time constraints impacted the creative task assigned to the participants. Prior research on the effects of intelligent assistant such as Alexa shows that interactions with this technology negatively affected users' performance (Edwards et. al., 2019). Furthermore, smartphones have also been to be a source of distraction during task completion which leads to lowered performance (Thornton et. al., 2014). In this study, we found a main effect for intelligent assistants such that teams with the device present performed the worst on a creative task. Here, it is also important to re-think the nature of the task assigned to team in our sample: the planning of a Sweet 16 birthday party. It can be reasonably assumed that most of our research participants must have been to one birthday party or more. Figuratively speaking, our intelligent assistant 'Alexa', has yet to be a guest at one. Yet, our study participants tended to rely on the intelligent to give creative ideas especially under time scarcity condition. One might argue, that it is not naive to assume that Alexa may have more ideas than a human being concerning a birthday, since it is an intelligent agent running on complex algorithms. To that we suggest that even if Alexa may "list" some ideas for a cake, it cannot provide combinations of flavors that can have a creative edge. Our findings demonstrate that the teams within a time scarce condition fixate on using Alexa through their allotted time for the task. This kind of team behavior is an exemplar of their bounded rationality. However, teams with more time at their disposal to learn the technology's shortcomings very quickly and their interaction with Alexa reduces as time passes (see Figure 1). These findings point to the direction of recent research where participants chose to prefer to rely on advice of algorithms and machines over people (Logg, Minson, & Moore, 2018; Robinette et. al., 2016).

That said, a close analysis demonstrates that scarce time can also negatively impact creativity, which is theoretically in line with previous findings (Amabile et. al., 2002). The results of this study uphold findings within the literature examining how time pressure can negatively affect a team's ability to perform well on creative tasks (Baer & Oldham, 2006). We echo findings from Kelly and Karau (1991) that demonstrate how under varying degrees of time constraints, individuals reduce collaborative interactive strategies in their teamwork, thus resulting in detrimental effects to creativity processes. These findings point toward potentially serious consequences where intelligent machines are being incorporated in human teams.

The purpose of studying the use of an intelligent assistant under time scarcity is to focus on decision-making in its descriptive sense i.e. what happens or what people do in the presence of an intelligent agent and time availability. Our goal is not to debate intelligent assistants from a normative point of view—a process which would require a judgement if this technology is good or bad for us (see Bell, 1988, for normative vs descriptive approaches to decision-making). In fact, this study is only meant to provide a nuanced understanding of how the deployment of intelligent assistants under environmental constraints such as time scarcity can impact the teams' interactions, technology use, and performance.

## 5.1 Limitations & Future Research

This study has some limitations; however, they can help pave the way for future research. First, this study recruited participants who have had no exposure to physical/smart speaker-type intelligent assistant from manufacturers such as Amazon, Microsoft or Google. As a first study, it was done to equalize participants on experience and remove any confounds that prior communication or use with an intelligent assistant could have created. However, it is plausible that more experienced participants can use such technologies to their advantage in the future in many ways. Second, our findings use dyads within the context of a creative task. Again, it is likely that different task types and group sizes will yield different results with the presence of intelligent assistant and varying time availability. Much remains to be explored and we suggest using various kinds of tasks types (see Edwards et. al., 2019; McGrath, 1984; McGrath, 1991) and group sizes as a starting point for further research on AI-enabled technology and its impact on team behaviors. Third, users' prior task and relationships with team members, and transactive memory systems within teams could impact the use of an intelligent assistant and subsequent decision-making. Fourth, this is a lab study which used an experimental method to control for as many variables as possible. However, other kinds research methods and contexts can produce differing and novel findings. Fifth, it is also important to justify that why we do not discuss the concept of optimal time for completing a task (c.f. Karau & Kelly, 1991). Optimality implies normativity (i.e., there is an "ideal" or "best" amount of time for this creative task). However, that is not the focus of this study. Further, methodological concerns prevent us from investigating such a perspective at this point of the study. The goal of the study is how intelligent assistant and time impact interactions, creativity and decision-making— and not how an intelligent assistant impacts optimality. Sixth, a lack of interaction between teammates who had an intelligent assistant available to them may also be accounted by the delay device may have in responding back. However, this by itself, has implications concerning the use of machines in teamwork. Finally, the analysis and results of this paper mostly rely on behavioral data. In our case, we argue that unobtrusive behavioral data provide stronger evidence than self-reported perceptions of technology use and team interactions. However, given various conditions and research questions; this may not be feasible and therefore, we encourage future researchers to explore other methodological arenas.

## 6. CONCLUSION

Human-AI collaboration--i.e., the implementation of AI-enabled technologies within human workspaces and teams--is being described as the future of work. However, there is a serious need to understand how these technological advances are impacting our work and interactions. Evidence from our study suggests that the deployment of an intelligent assistant in teams increases its use but decreases face-to-face verbal interactions between members, and creative performance under time scarcity. These findings have serious implications for the future of work. We hope that this study helps researchers, users, and developers in the area of human-AI collaboration and human-machine interaction and encourage a long line of research via theoretical arguments and empirical findings.